\newif\ifcheckpagelimits
\checkpagelimitstrue
\checkpagelimitsfalse

\ifcheckpagelimits
 \documentclass[nofootinbib,prd,aps,twocolumn,showpacs,showkeys,superscriptaddress,final,reprint,floatfix]{revtex4-1}
 \newcommand{\todo}[1]{}
\else
 \documentclass[prd,aps,twocolumn,showpacs,showkeys,superscriptaddress,final,reprint,floatfix]{revtex4-1}
 \newcommand{\todo}[1]{{\pdfmargincomment[icon=Note,color=pink]{#1}}}
\fi

\usepackage{mathptmx}
\usepackage{amssymb}
\usepackage{amsmath}
\usepackage{amsfonts}
\usepackage{array}

\usepackage{bm}
\usepackage{xcolor}
\usepackage{graphicx}

\usepackage{braket}

\definecolor{mygrey}{gray}{0.35}
\definecolor{myblue}{rgb}{0.2,0.2,0.8}
\definecolor{myzard}{cmyk}{0,0,0.05,0}
\definecolor{mywhite}{rgb}{1,1,1}
\definecolor{myred}{rgb}{1,0.,0.3}

\usepackage[colorlinks=true,citecolor=myblue,linkcolor=myred]{hyperref}

 \def\ee{\mathord{\rm e}}
 
 \def\ii{\mathord{\rm i}}

\def\half{\textstyle\frac{1}{2}}

\def\fourth{\textstyle\frac{1}{4}}

\DeclareMathOperator{\sign}{sign}

\renewcommand{\ii}{{\rm i}}
\renewcommand{\ee}{{\rm e}}
\def\beq{\begin{equation}}
\def\eeq{\end{equation}}

\def\barray{\begin{eqnarray}}
\def\earray{\end{eqnarray}}

\begin{document}

\title{Symmetry-protected Topological Phases in Lattice Gauge Theories: Topological QED$_2$  }

\author{G. Magnifico}
\affiliation{Dipartimento di Fisica e Astronomia dell'Universit\`a di Bologna, I-40127 Bologna, Italy}
\affiliation{INFN, Sezione di Bologna, I-40127 Bologna, Italy}

\author{D. Vodola}
\affiliation{Department of Physics, College of Science, Swansea University, Singleton Park, Swansea SA2 8PP, United Kingdom}

\author{E. Ercolessi}
\affiliation{Dipartimento di Fisica e Astronomia dell'Universit\`a di Bologna, I-40127 Bologna, Italy}
\affiliation{INFN, Sezione di Bologna, I-40127 Bologna, Italy}

\author{S. P. Kumar}
\affiliation{Department of Physics, College of Science, Swansea University, Singleton Park, Swansea SA2 8PP, United Kingdom}

\author{M. M\"{u}ller}
\affiliation{Department of Physics, College of Science, Swansea University, Singleton Park, Swansea SA2 8PP, United Kingdom}

\author{A. Bermudez}
\affiliation{Departamento de F\'isica Te\'orica, Universidad Complutense, 28040 Madrid, Spain}

\begin{abstract}
The interplay of symmetry, topology, and many-body effects in the classification of  phases of matter poses a formidable challenge  in condensed-matter physics. Such many-body effects are typically induced by inter-particle interactions  involving an action at a distance, such as the Coulomb interaction between electrons in a symmetry-protected topological (SPT) phase. In this work we show that similar phenomena also occur in certain relativistic theories with interactions mediated by gauge bosons, and constrained by gauge symmetry. In particular, we introduce a variant of the Schwinger model or quantum electrodynamics (QED) in 1+1 dimensions on an interval, which displays dynamical edge states localized on the boundary. We show that the system hosts SPT phases with a dynamical contribution to the  vacuum $\theta$-angle from edge states, leading to a new type of  {\em topological} QED in 1+1 dimensions. The resulting system displays an SPT phase which can be viewed as a correlated version of the Su-Schrieffer-Heeger topological insulator for polyacetylene due to non-zero gauge couplings. We use bosonization and density-matrix renormalization group techniques to reveal the detailed phase diagram, which can further be explored in experiments of ultra-cold atoms in optical lattices.

\end{abstract}

\maketitle

Global and local symmetries play a crucial role in our understanding  of Nature at  very different energy scales~\cite{qft_book,cm_book}. At high energies, they govern the behavior of fundamental particles~\cite{YM_non_abelian}, their spectrum and interactions~\cite{goldstone, higgs_1,*higgs_2}. At low energies~\cite{anderson_sc}, spontaneous symmetry breaking and local order parameters characterize a wide range of phases of matter~\cite{landau_sb} and a rich variety of collective phenomena~\cite{anderson_emergence}. There are, however, fundamental physical phenomena that can only  be characterized  by non-local order parameters, such as the Wilson loops distinguishing  confined and deconfined phases in gauge theories~\cite{kogut_review},  or  hidden order parameters distinguishing topological phases in solids~\cite{wen_book}. The former, requiring a non-perturbative approach to quantum field theory (e.g. lattice gauge theories (LGTs)), and the latter, demanding the introduction  of mathematical tools of topology in condensed matter (e.g. topological invariants), lie at the forefront of  research in  both high-energy and condensed-matter physics.

The interplay of symmetry and topology can lead to a very rich, and yet partially-uncharted, territory. For instance, different phases of matter can arise without any symmetry breaking:  {\it symmetry-protected topological} (SPT) phases. Beyond the celebrated integer quantum Hall effect~\cite{iqhe_exp,qhe_invariant,haldane_spt,kane_mele_z2}, a variety of SPT phases have already been identified~\cite{kane_rmp_ti,qi_rmp_ti,tis_exp} and  realized~\cite{pion_exps_1,*pion_exps_2}. Let us note that  some representative models of these SPT phases~\cite{ti_qft_representatives1, *ti_qft_representatives2} can be understood as lower-dimensional versions of the so-called domain-wall fermions~\cite{domain_wall_fermions}, introduced in the context of chiral symmetry in lattice field theories~\cite{chiral_lattice_review}. A current problem of considerable  interest is to understand strong-correlation effects in SPT phases as interactions are included~\cite{corr_ti_1, *corr_ti_2}, which may, for instance, lead  to exotic fractional excitations~\cite{fqhe_exp,fqhe_laughlin}. So far, the typical interactions considered  involve an action at a distance (e.g. screened Coulomb or Hubbard-like nearest or next-to-nearest neighbor interactions). To the best of our knowledge, and with the recent  exception~\cite{spt_gauge}, the study of  correlated SPT phases with mediated interactions remains a  largely-unexplored subject.


\begin{figure}
\centering
  \includegraphics[width=1\columnwidth]{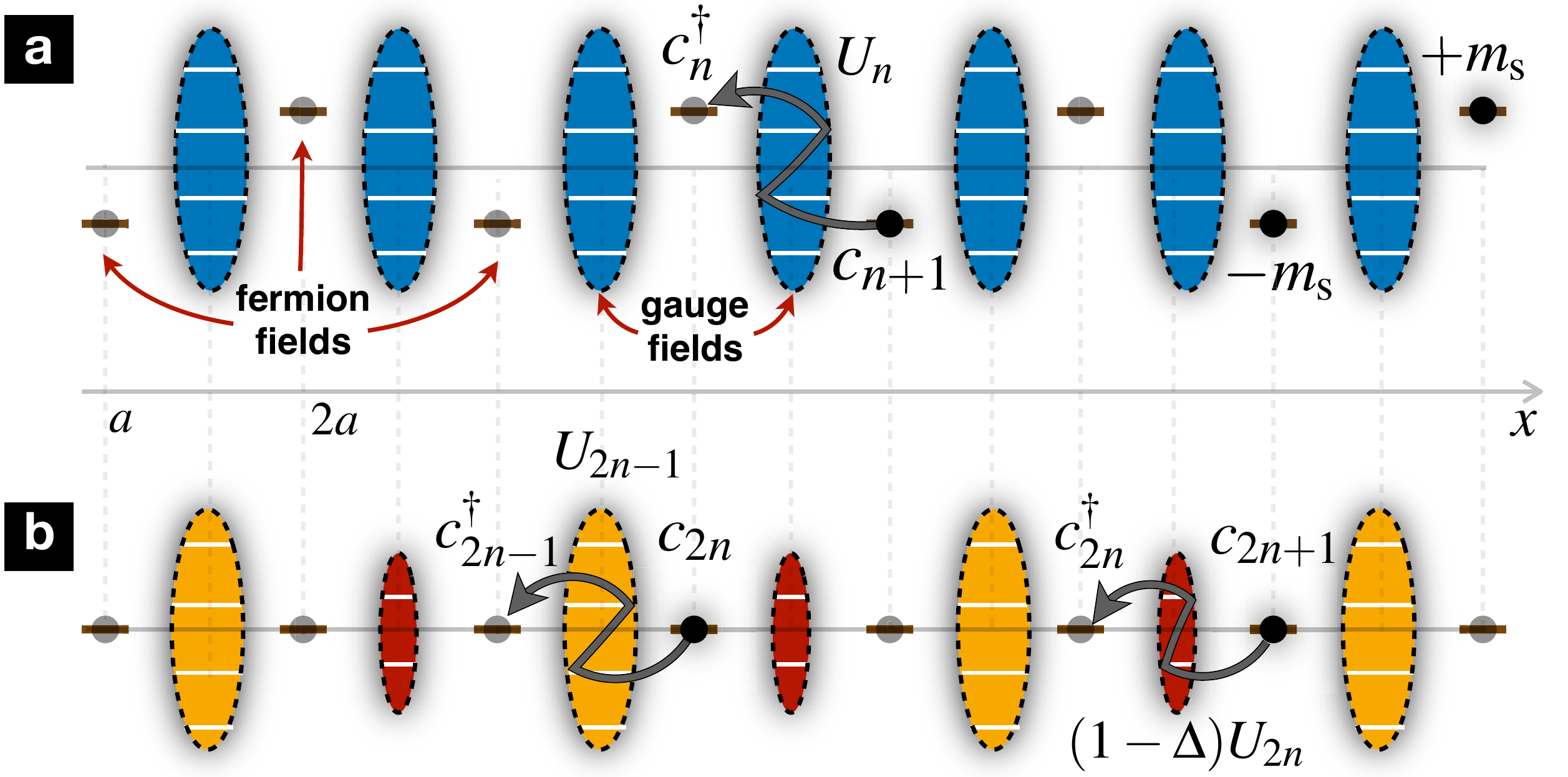}
  \caption{\label{Fig:discretization_scheme}  {\bf Discretizations for standard and topological QED$_2$}: {\bf (a)} Staggered-fermion approach to the massive Schwinger model. The relativistic Dirac field is discretized into spinless lattice fermions subjected to a staggered on-site energy $\pm m_{\rm s}$, represented by filled/empty circles in a 1D chain with alternating heights. The gauge field is discretized into rotor-angle operators that reside on the links, depicted as shaded ellipses with various levels representing the electric flux eigenbasis. The gauge-invariant term $c_n^\dagger U_n c_{n+1}$ involves the tunneling of neighboring fermions, dressed by a local excitation of the gauge field  in the electric-flux basis $U_n\ket{\ell}=\ket{\ell+1}$, represented by the zig-zag grey arrow joining two neighboring fermion sites, via an excitation of the link electric-flux level.  {\bf (b)} Dimerized-tunneling approach to the topological Schwinger model. The previous staggered mass is substituted by a gauge-invariant tunneling with alternating strengths $(1-\delta_n)c_n^\dagger U_n c_{n+1}$, where $\delta_n=0,\Delta$ for even/odd sites. This  dimerization of the tunneling matrix elements is represented by alternating big/small ellipses at the odd/even links.}
\end{figure}

In this work, we initiate a systematic study of SPT phases with interactions dictated by gauge symmetries focusing on the lattice Schwinger model, an Abelian LGT that regularizes quantum electrodynamics in 1+1 dimensions (QED$_2$)~\cite{schwinger_model}. We show that a discretization alternative to the standard lattice approach~\cite{ham_lgt} leads to a \emph{topological} Schwinger model, and derive its continuum limit  referred to as {\it topological QED$_2$}. This continuum quantum field theory  is used  to predict a phase diagram that includes SPT, confined, and  fermion-condensate phases, which are then discussed in the context of the aforementioned domain-wall fermions in LGTs.

 We benchmark our predictions based on bosonization techniques  against exhaustive  numerical simulations via  the density-matrix renormalization group (DMRG). 
Our study shows that SPT phases  also appear in  gauge theories, and it is conceivable that they will lead to a rich playground where topological effects coexist with non-perturbative  phenomena such as confinement and charge shielding, or string breaking~\cite{Shifman2012}. We note also that our study can be relevant for experimental realizations far from the high-energy-physics domain, as some of the discretized LGTs can be simulated by state-of-the-art experiments with cold atoms~\cite{QS_cold_atoms, QS_trapped_ions, qs_goals}. In particular, previous schemes of Bose-Fermi mixtures can be adapted to the quantum simulation of topological QED$_2$~\cite{new_topological_QED2}.

\emph{The Schwinger model}~\cite{schwinger_model} describes a Dirac fermion field $\Psi(x)$  of mass $m$ coupled to an electromagnetic field $A^{\mu}(x)$  in a (1+1)-dimensional Minkowski spacetime with coordinates $x^\mu=(t,{\rm x})$,  $\mu\in\{0,1\}$, and metric $\eta={\rm diag}(1,-1)$. After setting  $\hbar=c=1$, the Lagrangian density of the massive Schwinger model ($m{\rm S}$) is 
\beq
\label{eq:schwinger_lagrangian}
\mathcal{L}_{m{\rm S}}=\overline{\Psi}\left[\ii\gamma^\mu\left(\partial_\mu+\ii gA_\mu\right)-m\right]{\Psi}-\fourth F^{\mu\nu}F_{\mu\nu},
\eeq
where $\partial_\mu = {\partial}/{\partial x^{\mu}}$ and $A_\mu(x)=\eta_{\mu\nu}A^\nu(x)$. We have  introduced the adjoint $\overline{\Psi}(x)=\Psi^{\dagger}\!(x) \gamma^0$, the Dirac matrices $\gamma^\mu$ satisfying $\{\gamma^\mu,\gamma^\nu\}=2\eta^{\mu\nu}$, and the (bare) coupling  $g$ of the fermion current to the gauge field with  electromagnetic field tensor $F_{\mu\nu}=\partial_{\mu}A_{\nu}-\partial_{\nu}A_{\mu}$. The physics of $\mathcal{L}_{m{\rm S}}$ is periodic in  the so-called vacuum~$\theta$ angle, a term proportional to the background electric field~\cite{schwinger_theta_vacuum, schwinger_theta_chiral}. In the massive case, a continuous phase transition between  confined  and  symmetry-broken phases with a fermion condensate occurs for $\theta=\pi$~\cite{massive_schwinger_theta}. Moreover, the Schwinger model captures some of the significant non-perturbative effects of higher-dimensional   non-Abelian gauge theories mentioned above~\cite{massive_schwinger_shielding, schwinger_anomaly, massive_schwinger_shielding, string_breaking}.  

Various numerical techniques~\cite{schwinger_mps_review}, including finite-lattice methods~\cite{lattice_schwinger_critical_point},  exact diagonalization~\cite{schwinger_exact_diagonalization}, Monte Carlo~\cite{schwinger_monte_carlo},  DMRG~\cite{schwinger_dmrg} and matrix-product states~\cite{schwinger_mps_1,*schwinger_mps_2, *schwinger_mps_3}, have been used to  unveil this  non-perturbative phenomenology. These methods typically rely on the Kogut-Susskind discretization~\cite{ham_lgt}  [see Fig.~\ref{Fig:discretization_scheme}{\bf (a)}], where: {\it (i)} the spatial coordinates are discretized into  a chain ${\rm x}=n a$ of   lattice spacing $a$, where  $n\in\mathbb{Z}_{N_{\rm s}}$ labels the   number of  sites $N_{\rm s}$; {\it (ii)} the fields $\Psi(x)$ are represented by lattice fermions $c_n$ with an alternating staggered mass $m_{\rm s}$;  {\it (iii)} the gauge field sector is represented by rotor-angle operators $L_n,\Theta_n$  assigned to the links at  ${\rm x}=(n+\half) a$. The angle operator is related to the gauge field $\Theta_n=agA_1({\rm x})$, while the rotor corresponds to an angular-momentum operator related to the  electric field $L_n=E({\rm x})/g=F_{01}({\rm x})/g$, which is diagonal in  the basis $\ket{\ell}$, i.e. $L_n\ket{\ell}=\ell\ket{\ell}$ for $\ell\in\mathbb{Z}$. In this way, the LGT Hamiltonian for the  massive Schwinger model becomes
\beq
\label{eq:standard_KS_schwinger}
H_{m{\rm S}}\!=a\!\sum_{n=1}^{N_{\rm s}} \!\left(\!\frac{-1}{2a}\!\left(\ii c_n^{{\dagger}}U_{n}^{\phantom{\dagger}}c_{n+1}^{\phantom{\dagger}}+{\rm H.c.}\!\right)\!+m_{\rm s}(-1)^nc_n^{{\dagger}}c_n^{\phantom{\dagger}}+\frac{g^2}{2}L_n^2\right)\!.
\eeq
Here, we have introduced  the  link operators $U_{n}=\ee^{\ii\Theta_n}$, which  act as unitary ladder operators $U_{n}\ket{\ell}=\ket{\ell+1}$. In the continuum limit $a\to 0$, one recovers the Hamiltonian quantum field theory associated to Eq.~\eqref{eq:schwinger_lagrangian} with a Dirac mass $m=m_{\rm s}$~\cite{ham_lgt} .

In this work, we introduce an alternative discretization, which not only reproduces Eq.~\eqref{eq:schwinger_lagrangian} in the continuum limit, but   also hosts  an  SPT phase where the fermions interact via the gauge field. Note  that the discretized model~\eqref{eq:standard_KS_schwinger} has a two-site unit cell, as the staggered mass breaks explicitly the lattice translational invariance.  An alternative discretization that maintains this property follows from the dimerization of the tunnelings with a two-site periodicity [see Fig.~\ref{Fig:discretization_scheme}{\bf (b)}], yielding the {\it topological lattice Schwinger model}
\beq
\label{eq:dimerised_KS_schwinger}
H_{t{\rm S}}\!=a\!\sum_{n=1}^{N_{\rm s}} \!\left(\!\frac{-1}{a} \left(\ii (1-\delta_n)c_n^{{\dagger}}U_{n}^{\phantom{\dagger}}c_{n+1}^{\phantom{\dagger}}+{\rm H.c.}\!\right)\!+\frac{g^2}{2}L_n^2\right)\!,
\eeq
where the dimerization satifies $\delta_{2n}=0$ and $\delta_{2n-1}=\Delta$. 

\emph{Topology in the continuum limit}.-- It is  now natural to ask if the continuum limit of Eq.~\eqref{eq:dimerised_KS_schwinger} indeed contains  the Hamiltonian of the massive Schwinger model~\eqref{eq:schwinger_lagrangian}. To this end, let us  first set $g = 0$, such that  $H_{t{\rm S}}=H_{\rm SSH}+a\sum_nE({\rm x})^2/2$, where $H_{\rm SSH}$ corresponds  to the Su-Schrieffer-Heeger (SSH) model of polyacetylene in the limit of a static lattice~\cite{ssh_polyacetylene,polyacetylene_rmp}, a paradigmatic example of an SPT Hamiltonian~\cite{table_ti_1, *table_ti_2} displaying edge states for $\Delta \in (0,2)$. To find the continuum limit of the interacting $H_{t{\rm S}}$\eqref{eq:dimerised_KS_schwinger},   the existence of these edge states must be taken  into account (see Appendix~\ref{section:supp_bosonization}). In particular, for  $0<\Delta\ll 1$,  one finds $H_{\rm SSH}=\int_{0}^{aN_{\rm s}} {\rm dx}\mathcal{H}_{t{\rm D}}$, where 
 \beq
 \label{eq:correct_continuum_ssh}
 \mathcal{H}_{t{\rm D}}=\overline{\Psi}({\rm x})\left(-\ii\gamma^1\partial_{\rm x}+\frac{\Delta}{a}\right){\Psi}({\rm x})+\sum_{\eta={\rm L,R}}\!\!\!\epsilon_{\eta}|\chi_{\eta}({\rm x})|^2\hat{\eta}^\dagger\hat{\eta}.
 \eeq
Here, in addition to the bulk Dirac fermions of mass $\Delta/a$, we have also included the left $\hat{\mathrm{L}}$ and right $\hat{\mathrm{R}}$ topological edge states   with energy $\epsilon_{\mathrm{L},\mathrm{R}}$. These states have wave functions 
\beq
\label{eq:edge_states}
{\chi}_{\rm L}({\rm x}) \sim \ee^{-\frac{{\rm x}}{\xi}}\sin(k_{\rm F} {\rm x}),\hspace{1ex}{\chi}_{\rm R}({\rm x})\sim \ee^{-\frac{(L_s-{\rm x})}{\xi}}\sin(k_{\rm F} (L_s-{\rm x}))
\eeq
where $\xi=a/\Delta$ is a localization length of the exponential decay, $L_s = aN_{\rm s}$ is the length of the system, and   $k_{\rm F} = \pi/2 a$ is the momentum around which the continuum limit is computed. Let us note that these edge states can be interpreted as lower-dimensional versions of the domain-wall fermions~\cite{domain_wall_fermions}, which becomes apparent~\cite{wilson_gross_neveu} after connecting the SSH-type discretization~\cite{ssh_polyacetylene} to the Wilson-type approach~\cite{wilson_fermions}.

The Hamiltonian of Eq.~\eqref{eq:correct_continuum_ssh} forms the matter sector of the topological Schwinger model~\eqref{eq:dimerised_KS_schwinger}, which in the Coulomb gauge $A_1=0$ becomes $H_{t{\rm S}}=\int {\rm dx}\mathcal{H}_{t{\rm S}}$, where 
\beq
\begin{split}
\label{eq:continuum_ts}
 \mathcal{H}_{t{\rm S}}&=  \mathcal{H}_{t{\rm D}} -g A_0({\rm x})\overline{\Psi}({\rm x})\gamma^0{\Psi}({\rm x})+\fourth F^{\mu\nu}({\rm x})F_{\mu\nu}({\rm x}).
 \end{split}
 \eeq
The gauge field theory~\eqref{eq:correct_continuum_ssh}-\eqref{eq:continuum_ts} is a new type of {\it topological QED$_2$}  describing the interaction of the bulk relativistic fermions and the topological  edge modes with the gauge field,  according to  the $U(1)$ local symmetry characteristic of QED. 

\emph{Bosonization analysis}.-- Bosonization has been used~\cite{massive_schwinger_shielding,massive_schwinger_theta,schwinger_theta_chiral} to prove that the massless Schwinger model is described by a Klein-Gordon field theory of mass $\mu=g/\sqrt{\pi}$. Here we apply bosonization  to obtain quantitative results about the phase diagram of Eq.~\eqref{eq:continuum_ts}, unveiling an interesting interplay of the edge states and the vacuum $\theta$ angle discussed above.

The bosonization dictionary relating fermionic fields $\Psi({\rm x}), \overline{\Psi}({\rm x})$ to bosonic ones $\phi({\rm x}),\Pi(\rm x)$ is given by
\beq
\label{eq:bosonization_identities}
\begin{split}
-\ii:\overline{\Psi}({\rm x})\gamma^1\partial_{{\rm x}}\Psi({\rm  x})\!:_\Delta\hspace{1ex}\xrightarrow{\hspace*{0.3cm}} \hspace{1ex}&:\half \Pi^2({\rm x})+\half(\partial_{{\rm x}}\phi({\rm x}))^2\!:_\mu,\\
:\overline{\Psi}({\rm x})\Psi({\rm x})\!:_\Delta\hspace{2ex}\xrightarrow{\hspace*{1.2cm}}\hspace{1ex} &-c\mu:\cos\left(2\sqrt{\pi}\phi({\rm x})\right)\!:_\mu,\\
:{\Psi}^\dagger({\rm x})\Psi({\rm x})\!:_\Delta \hspace{1ex}\xrightarrow{\hspace*{1.2cm}}\hspace{1ex}&\partial_{\rm x}\phi/\sqrt{\pi}.\\
\end{split}
\eeq
where $c=\ee^\gamma/2\pi$ with  Euler's constant  $\gamma\approx0.5774$, and $:\!(\hspace{1ex})\!:_m$ denotes normal ordering of the Fermi or Bose fields with respect to the  fermion (boson) mass $m=\Delta/a$ ($\mu=g/\sqrt{\pi}$).  The first two  relations can be used to transform the matter sector of Eq.~\eqref{eq:continuum_ts}. The last expression can be used, in combination with Gauss' law $\partial_{\rm x}E_{\rm bulk}({\rm x})=g:{\Psi}^\dagger({\rm x})\Psi({\rm x}):_\Delta$, to bosonize also the gauge-field contribution. This  leads  directly to $ E_{\rm bulk}({\rm x})={g}\left(\phi({\rm x})+\frac{\theta}{2\sqrt{\pi}}\right)/{\sqrt{\pi}}$,  where one sees how the vacuum angle $\theta=2\pi E_{\rm ext}/g$ originates from a constant  field  after the integration of Gauss' law.

The novel ingredient for the bosonization of topological QED$_2$~\eqref{eq:continuum_ts} is to consider that the Gauss' law must be modified as, in the SPT phase, the  edge states can also contain charges. Focusing on the regime $0<\Delta\ll 1$, where the edge-state localization length  is very small, one can consider that the boundary charge only penetrates   into a  small region  close to the edges. Considering  the boundary conditions for the electromagnetic field across this region, which imply that the normal component of the electric field must be discontinuous, we find that  
$ E_{\rm edge}({\rm x})=g\left(\sign(x)\hat{\rm L}^\dagger\hat{\rm L}+\sign(x-L_s)\hat{\rm R}^\dagger\hat{\rm R}\right)/2$. Essentially, the  regions  that contain a charge contribute with a constant electric field  of $+g/2$ to its right and $-g/2$ to its left, as is known already for 1D classical electrodynamics~\cite{byrnes_phd}.

Substituting $E({\rm x})= E_{\rm bulk}({\rm x})+ E_{\rm edge}({\rm x})$ with the bosonization identities into $\mathcal{H}_{t{\rm S}}$ we find
\beq
\label{eq:bosonized_schwinger_model}
\begin{split}
\mathcal{H}_{t{\rm S}}&=\sum_{\eta}\epsilon_{\eta}|\chi_{\eta}({\rm x})|^2\hat{\eta}^\dagger\hat{\eta}+\half \Pi^2({\rm x})+\half(\partial_{{\rm x}}\phi({\rm x}))^2\\
&+\frac{g^2}{2\pi}\left(\phi({\rm x})+\frac{1}{2\sqrt{\pi}}\hat{\theta}\right)^2 -c\mu\frac{\Delta}{a}\cos\left(2\sqrt{\pi}\phi({\rm x})\right),
\end{split}
\eeq
 where normal ordering with respect to the mass $\mu$ is assumed. Here, the vacuum angle has turned into a dynamical operator 
\beq
\label{eq:theta_angle_operator}
\hat{\theta}=\theta+\pi\left(\sign(x)\hat{\rm L}^\dagger\hat{\rm L}-\sign(x-L_s)\hat{\rm R}^\dagger\hat{\rm R}\right). 
\eeq
Equations~\eqref{eq:bosonized_schwinger_model} and~\eqref{eq:theta_angle_operator} are the main result of this work: they show that the new vacuum angle is not simply a $c$-number, or an adiabatic classical field~\cite{axion_ti_1,*axion_ti_2}, but a quantum-mechanical operator depending on the constant external field via $\theta=2\pi E_{\rm ext}/g$, and  on the density of the topological edge states. 
Moreover, $\mathcal{H}_{t{\rm S}}$ incorporates the interplay of $\hat{\theta}$ with the 1D electromagnetic field, which is not an external field, but rather  obeys its own dynamics. As we  now show, the combination of these ingredients  leads to exotic effects in topological QED$_2$.

\begin{figure}
\centering
\includegraphics[width=0.95\columnwidth]{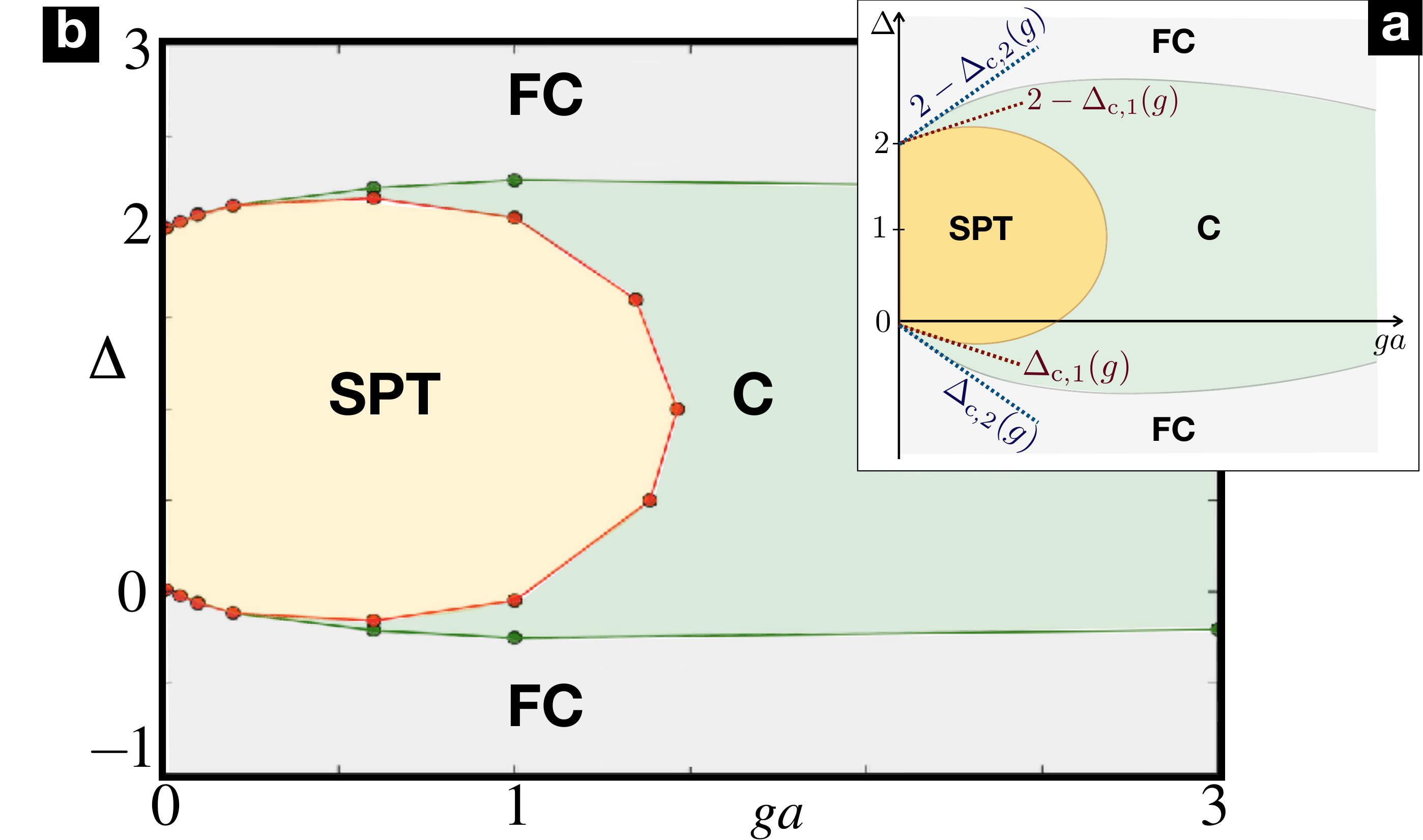}
\caption{\label{Fig:qualitative_phase_diagram} {\bf Phase diagram of topological QED$_2$:} {\bf (a)} obtained from the bosonized Hamiltonian \eqref{eq:bosonized_schwinger_model}. We predict three distinct phases: a symmetry-protected topological phase (SPT), a confined phase (C) and a symmetry-broken fermion condensate (FC); {\bf (b)} obtained by DMRG confirming the analytical phase diagram in {\bf (a)}.}
\end{figure}

\emph{Topological phase diagram}.-- Let us now discuss the phase diagram of  topological QED$_2$ at $\theta=\pi$ for generic $(\Delta, ga)$. 
To this end, we define an effective potential $V(\phi)$ by shifting the scalar field $\phi({\rm x})\to \phi({\rm x})- \hat{\theta}/2\sqrt{\pi}$ in~\eqref{eq:bosonized_schwinger_model}
\beq
 V(\phi)=\frac{g^2}{2\pi}\phi^2({\rm x})-c\mu\frac{\Delta}{a}\cos\left(2\sqrt{\pi}\phi({\rm x})-\hat{\theta}\right),
\eeq
and  treat it semiclassically for a small $g$. The potential $V(\phi)$ will make the non-interacting critical point $\Delta_{\rm c}=0=g_{\rm c}$ flow to other critical points depending on $g$, $\Delta$ and $\hat{\theta}$.

 For $0<\Delta\ll ga$ (in the SPT phase),  $\braket{\hat{\theta}}_{\rm gs}=0$ (mod $2\pi)$ and the cosine term in $V(\phi)$ will renormalize the mass of~$\phi$ as $\mu(g,\Delta)={g}\left(1+{2\sqrt{\pi} \ee^\gamma \Delta}/{ga}\right)^{1/2}/{\sqrt{\pi}}$ at  leading order. Thus,   a new critical line is found when $\mu(g,\Delta)$ vanishes, i.e. $\Delta_{\rm c,1}(g)=-{ga}\: \ee^{-\gamma}/{2\sqrt{\pi}}$ [red dashed line in Fig.~\ref{Fig:qualitative_phase_diagram}{\bf (a)}], such that the correlated SPT phase with gauge-field couplings extends to the region $\Delta\gtrsim \Delta_{\rm c,1}(g)$  . When $\Delta\lesssim \Delta_{\rm c,1}(g)$ (out  of the SPT phase),  $\braket{\hat{\theta}}_{\rm gs}=\pi$ and the quadratic term in $V(\phi)$ dominates yielding a ground-state with $\langle \phi\rangle_{\rm gs}=0$. This is a {\it confined phase} [C in Fig.~\ref{Fig:qualitative_phase_diagram}{\bf (a)}] displaying {\it fermion trapping}, as the spectrum only shows  massive bosonic excitations  understood as mesons, i.e. strongly bound fermion-antifermion pairs~\cite{massive_schwinger_theta}. On the other hand, when $\Delta\ll \Delta_{\rm c,1}(g)$, the cosine in $V(\phi)$ dominates, yielding  a ground-state with $\langle \phi\rangle_{\rm gs}\neq 0$ that spontaneously breaks the $\mathbb{Z}_2$ symmetry $\phi({\rm x})\to-\phi({\rm x})$. This phase is a  {\it fermion condensate} [FC in Fig.~\ref{Fig:qualitative_phase_diagram}{\bf (a)}] as it displays both $\langle  E({\rm x})\rangle \neq 0$ and $\langle  \overline{\Psi}({\rm x})\ii\gamma^5{\Psi}({\rm x})\rangle\neq 0$~\cite{schwinger_dmrg}. The C-FC phase transition must be  analogous to the one in the standard massive Schwinger model for $\theta=\pi$~\eqref{eq:standard_KS_schwinger}. Using the results of this well-studied model~\cite{schwinger_dmrg}, we conjecture that the second critical line is $\Delta_{\rm c,2}(g)=-ga/3$ [dashed blue line in Fig.~\ref{Fig:qualitative_phase_diagram}{\bf (a)}]. 

Using these bosonization predictions and considering that, at very strong couplings, the SPT phase disappears in favor of the confined phase, we draw the qualitative phase diagram of topological QED$_2$ in Fig.~\ref{Fig:qualitative_phase_diagram}{\bf (a)}, taking into account the symmetry around $\Delta=1$. Let us now discuss this phase diagram in the context of domain-wall fermions in LGTs~\cite{domain_wall_fermions,dw_fermions_various1,*dw_fermions_various2,*dw_fermions_various3,*dw_fermions_various4}. As advanced above, the connection~\cite{wilson_gross_neveu}  of the SSH-type discretization to a Wilson-type approach~\cite{wilson_fermions} indicates that the above SPT phase corresponds to the parameter region where domain-wall fermions are expected~\cite{chiral_lattice_review}. The bosonization phase diagram is qualitatively similar to the phase diagram in lattice field theories with Wilson fermions~\cite{aoki}. In such theories, fermion condensates that spontaneously break the parity symmetry are known as Aoki phases, and are believed to be mere lattice artefacts due to the Wilson-type discretization. Note that, in our model, the parity-broken fermion condensate is not a discretization artefact, as it also appears in the  continuum limits of the standard Schwinger model~\cite{massive_schwinger_theta} and, as argued above, in  our topological Schwinger model.

\begin{figure}
\centering
  \includegraphics[width=0.95\columnwidth]{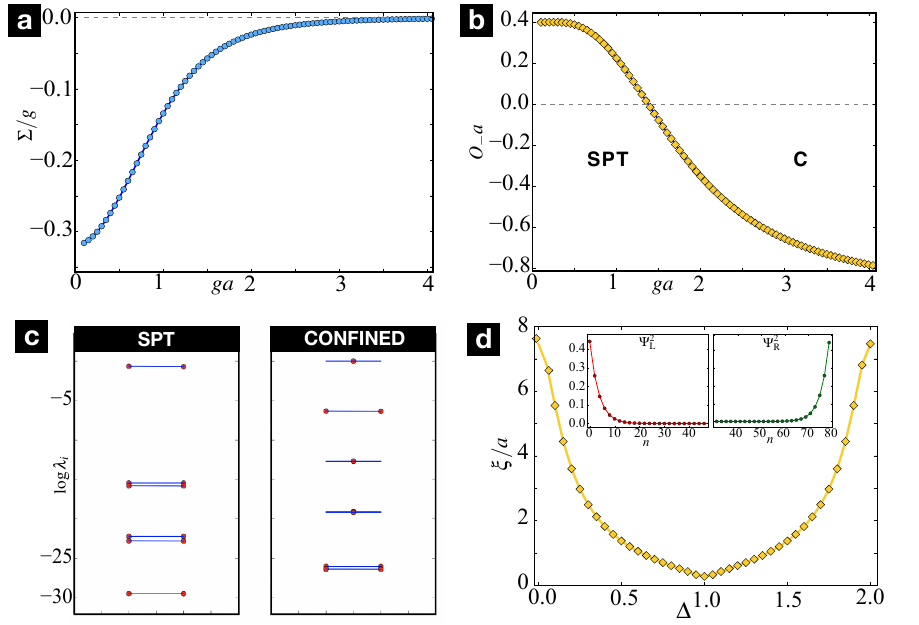}
  \caption{\label{Fig:o_param_1} {\bf (a)} Electric field order parameter $\Sigma$ as a function of $g$, for $\Delta = 0.5$ and $N_{\rm s}=80$. {\bf (b)} Topological order parameter $O_{-}$ for the same parameters as {\bf (a)}. {\bf (c)} Entanglement spectrum for $\Delta=0.5$ and (left) $g=0.2$ within the SPT phase, showing an accurate double degeneracy, and (right) $g = 5.0$ within the confined trivial phase. {\bf (d)} Many-body edge states: (main panel) Localization length $\xi$ for the zero-energy edge modes as a function of $\Delta$. (insets) Probability density  of the left and right-most edge modes ($\Delta=0.5, ga=0.2$).}
\end{figure}

\emph{DMRG analysis}.-- We explore the properties of the ground state $\ket{\text{gs}}$ of Eq.~\eqref{eq:dimerised_KS_schwinger} via a DMRG algorithm~\cite{dmrg_white} that  uses an alternative to the U(1) gauge fields: the $Z_N$ approach~\cite{zn_presentation, zn_study} (see Appendix \ref{sec:ZnModelHamiltonian}). We will focus on the case $N=3$, while the details of $N>3$ will be discussed  elsewhere~\cite{new_topological_QED2}.  The phase diagram can be characterized by computing the {\it electric-field order parameter} $\Sigma = \sum_{n=1}^{N_{s}} \bra{\rm gs} E_{n}  \ket{\rm gs}/{N_{\rm s}}$, and a {\it topological order parameter}  (introduced for the SSH model in Ref.~\cite{ssh_order_parameters}) $O_{-} =2\sum_{n=1}^{N_{\rm s}/2} \bra{\rm gs}O^{(2n-1)}_{-}\ket{\rm gs}/N_{\rm s}$ with
\beq
\label{eq:magnifico_topological_order_quantity}
O^{(j)}_{-} = \frac{3}{2}\left( c_{j}^{\dagger}c_{j+1}^{\phantom{\dagger}} +  c_{j+1}^{\dagger}c_{j}^{\phantom{\dagger}}  \right) + a\rho_{j} \rho_{j+1} -  \frac{1}{2}(\rho_{j}+\rho_{j+1}),
\eeq
where  $\rho_{j}= c^\dagger_jc^{\phantom{\dagger}}_j $ is the number of fermions at each site.

If the electric field $\Sigma$ is positive (negative), the ground state is dominated by mesons (anti-mesons), while a positive (negative) $O_{-}$ signals a topologically non-trivial (trivial) phase. Figure~\ref{Fig:o_param_1} shows $\Sigma$ in panel \textbf{(a)} and $O_{-}$ in panel \textbf{(b)} as a function of  $g$, for a system with $N_{s}=80$ sites for $\Delta=0.5$. At small $g$,  the ground-state consists of a superposition of the anti-meson states (with negative electric field between couples) and the Dirac vacuum, which follows from  the standard interpretation of the Kogut-Susskind discretization. Moreover, the positive values for $O_{-}$ show that the system is in a non-trivial SPT phase. By increasing $g$, the ground state becomes eventually the topologically-trivial Dirac sea without electric or matter/antimatter excitations ($\Sigma \approx 0, O_{-} < 0 $). By a finite-size scaling analysis  of $\Sigma$ and $O_{-}$ (see Appendix \ref{sec:critical_lines_scaling}), we identify the critical points $(g_c, \Delta_c)$ of the transitions SPT-C and FC-C, and determine the complete phase diagram of the $\mathbb{Z}_3$ topological Schwinger model [Fig.~\ref{Fig:qualitative_phase_diagram}\textbf{(b)}], which is in perfect agreement with the analytical prediction previously described [Fig.~\ref{Fig:qualitative_phase_diagram}\textbf{(a)}].

In order to understand the robustness of the SPT phase for different $\mathbb{Z}_N$ algebras we also computed the critical point $g_c(N)$ related to the transition SPT-C on the line $\Delta=1$ for the $\mathbb{Z}_{5}$ and $\mathbb{Z}_{7}$ models. As reported in (see Appendix \ref{sec:robustness_other_n}), the critical value $g_{c}(N)$ grows with $N$ and approaches a finite value in the $N \rightarrow  \infty$ limit given by $g_{c}(\infty)a = 2.979$, showing that the SPT phase has a finite region of stability, in accordance to the previous analytical results.

We give further numerical evidence of the topological nature of the SPT phase by computing the {\it entanglement spectrum} and the {\it wave functions} of the  many-body zero-energy edge modes. The entanglement spectrum~\cite{ent_spectrum}  is defined as the set of the logarithm of the eigenvalues $\lambda_i$ of the reduced density matrix $\tilde{\rho}_{A}$ of one of the two complementary halves $A$ or $\bar{A}$ in which we partition the system. According to Refs.~\cite{entanglement_spectrum_topology, entanglement_spectrum_topology_2}, an even degeneracy of the entanglement spectrum is a hallmark for a non-trivial topological phase. As shown in the left panel of Fig.~\ref{Fig:o_param_1}\textbf{(c)}, for a small gauge coupling $g$, we find doublets in this spectrum, thus confirming that $\ket{\text{gs}}$ lies in an SPT phase. On the contrary, this degeneracy  disappears for a larger $g$ (right panel), which corresponds to the trivial confined phase. For the edge-modes {\it wave functions}, we follow (see Appendix \ref{sec:edge_mode_wf}) and we obtain Fig.~\ref{Fig:o_param_1}{\bf (d)}. The two insets show the squares of the left-most $\Psi_\text{L}$ and right-most $\Psi_\text{R}$ wave functions that, similar to Eq.~\eqref{eq:edge_states},  decay exponentially with the lattice site $n$ from the boundaries of the system. By fitting $\Psi_\text{L}$ and $\Psi_\text{R}$ with an exponential function we extract the localization length $\xi$ as a function of $\Delta$ [main panel of Fig.~\ref{Fig:o_param_1}{\bf (d)}]. $\xi$ is very small deep in the topological phase, while $\xi$ grows as we approach the critical points where the edge states delocalize into the bulk .

{\it Cold-atom quantum simulators}.-- As advanced previously, and detailed in~\cite{new_topological_QED2},  Bose-Fermi mixtures may allow for the realization of the topological Schwinger model~\eqref{eq:dimerised_KS_schwinger}. Let us now summarise the main ingredients of our scheme, starting from a previous proposal for the quantum simulation of the standard Schwinger model~\cite{qs_schwinger}.  We consider two-component fermionic atoms that are trapped in a 
blue-detuned tilted optical lattice, representing  the matter sector of the topological Schwinger model~\eqref{eq:dimerised_KS_schwinger},  while two-component bosons are confined by a much deeper
 red-detuned optical lattice, and will be employed to simulate the gauge field. As such, the bare tunneling of the bosons is inhibited, such that the $s$-wave scattering yields the electric energy of the lattice model~\eqref{eq:dimerised_KS_schwinger} by using the Schwinger-boson representation of the link operators~\cite{qs_schwinger}. For a sufficiently-large tilting, the bare fermionic tunnelling  is also inhibited, and one must only consider the fermion-fermion and fermion-boson $s$-wave scattering. Whereas the former are irrelevant for certain initial states, the latter must be exploited to achieve the gauge-invariant dimerized tunneling of Eq.~\eqref{eq:dimerised_KS_schwinger}. The possibility considered in our work is to use a state-dependent time-periodic modulation of the bosonic on-site energies, resonant with the fermionic tilting, to assist the fermion-boson spin-changing collisions from the $s$-wave scattering that exactly correspond to the gauge-invariant tunnelling. It can be shown that, under certain conditions~\cite{new_topological_QED2}, the dimerization of the  gauge-invariant tunnelling can be controlled via the modulation parameters, whereas spurious terms can be neglected or compensated. 

{\it Conclusions and outlook}.-- In this work, we have introduced an alternative discretization of the massive Schwinger model hosting a correlated SPT ground-state where the interactions between the fermions are mediated by gauge bosons. Using bosonization, we have shown that the underlying topology of the SPT phase upgrades the vacuum $\theta$ angle into a quantum operator that depends on the edge-state densities, and leads to a richer phase diagram in comparison to the standard Schwinger model. By DMRG simulations, we have carefully benchmarked the bosonization predictions by computing the complete phase diagram of the model and the relevant fingerprints of the correlated SPT phase, such as the entanglement spectrum and many-body edge states, finding perfect agreement with the bosonization results. 

The connection of these SPT phases to the so-called domain-wall fermions in LGTs point to an interesting avenue of research: the exploitation of topological features (e.g. degeneracies in the entanglement spectrum) to unveil the rich phase structure in LGTs. In this context,  the appearance of parity-breaking condensates in the continuum limit of our model contrast with the so-called Aoki phases in LGTs, which are typically considered as mere lattice artefacts not present in the continuum QFT. Another interesting difference with respect to the conjectured  phase diagram of non-Abelian LGTs displaying Aoki phases~\cite{aoki} is that our parity-broken condensate is not directly connected to the SPT region hosting domain-wall fermions. Our bosonization and DMRG results indicate that the confined phase, which preserves parity, extends all the way down to $g=0$, separating the SPT phase from the parity-broken fermion condensate. In the future, it would be very interesting to understand these connections/differences in more detail, and explore how the tools used to characterize  SPT phases might be useful for the understanding of QCD-like lattice theories in higher dimensions. Another interesting  topic is the study of dynamical effects that can evidence the interplay of topological features in SPT phases and non-perturbative  effects in LGTs.

\begin{acknowledgments}
E.E. and G.M. are partially supported by INFN through the project QUANTUM. S.P.K. acknowledges the support of  STFC grant ST/L000369/1. A.B.  acknowledges support from RYC-2016-20066,  FIS2015-70856-P,  and CAM regional research consortium QUITEMAD+. A.B. thanks T. Byrnes for kindly sharing the document of his PhD thesis~\cite{byrnes_phd},  A. Celi, P. Silvi and P. Zoller for interesting discussions, and  A. Celi for bringing the work~\cite{spt_gauge} to our attention.
\end{acknowledgments}

\appendix

\section{Topological QED\texorpdfstring{$_2$}{2}: non-interacting limit} 
\label{section:supp_bosonization}
In this Section we review the properties of the Su-Schrieffer-Heeger (SSH) model of polyacetylene~\cite{ssh_polyacetylene,polyacetylene_rmp} that corresponds to the discretized  non-interacting Schwinger model we introduced. We place a special emphasis to the connection to one-dimensional topological insulators, a paradigmatic example of an SPT phase and we also show how the non-trivial topological properties of the SSH model have to be considered to compute its continuum limit properly.

\subsection{SSH Model - SPT phase and topological invariant}
In the limit of vanishing coupling $g=0$, the Hamiltonian of the discretized Schwinger model  of Eq.~\eqref{eq:dimerised_KS_schwinger} of the main text reduces to $H_{t{\rm S}}=H_{\rm SSH}+a\sum_nE({\rm x})^2/2$, such that the matter sector decouples from the gauge-field sector and can be described by 
\beq
\label{eq:SSH}
H_{\rm SSH}=-\ii\sum_{n=1}^{N_s/2} (1-\Delta)a^\dagger_nb_n^{\phantom{\dagger}}+b^\dagger_na_{n+1}^{\phantom{\dagger}}+{\rm H.c.},
\eeq
where we have rewritten the even (odd) fermionic operators  $c_{2n}$ $(c_{2n-1})$ using a two-site unit cell notation $b_{n}$ $(a_{n})$. By performing a Fourier transform for  periodic boundary conditions, one obtains $H_{\rm SSH}=\sum_{k\in{\rm BZ}}\Psi^\dagger_kh(k)\Psi^{\phantom{\dagger}}_k$, where $h(k)=\boldsymbol{d}(k)\cdot\boldsymbol{\sigma}$ is the single-particle Hamiltonian, and  $\Psi^{\phantom{\dagger}}_k=(a_k,b_k)^{\rm t}$ is defined   within the first Brillouin zone ${\rm BZ}=[-\pi/a,\pi/a)$. In this expression, 
 $\boldsymbol{d}(k)=(-\sin ka, (1-\Delta-\cos ka),0)/a$, and $\boldsymbol{\sigma}$ is the vector of all three Pauli matrices $\boldsymbol{\sigma}=({\sigma}^x,{\sigma}^y,{\sigma}^z)$. 
 Note that the dimerization leads to a  momentum-dependent  mass $m_{t}(k)=(1-\Delta-\cos ka )/a$, a so-called  {\it topological mass} that plays a crucial role in the appearance of the SPT phase. 
 
A na\"{i}ve   long-wavelength approximation would yield  $H_{\rm SSH}=\int {\rm dx}\mathcal{H}_{m{\rm D}}$, where
 \beq
 \label{eq:naive_continuum_limnit_PBC}
 \mathcal{H}_{m{\rm D}}=\overline{\Psi}({\rm x})(-\ii\gamma^1\partial_{\rm x}+m){\Psi}({\rm x})
 \eeq
 is the Hamiltonian density for a massive Dirac field with $\gamma^0=\sigma^y$, $\gamma^1=\ii\sigma^z$,  and mass $m=-\Delta/a$ for  dimerizations $\Delta \ll 1$. 
 
 Here, we have introduced  the effective Dirac spinor $\Psi({\rm x})=(\psi^{\phantom{\dagger}}_u({\rm x}),\psi_d^{\phantom{\dagger}}({\rm x}))^{\rm t}$ for a small region  around the origin of the Brillouin zone $|k|<\Lambda_{\rm c}$ with components defined by
 \beq
 \label{eq:ssh_low_energy}
\psi_u^{\phantom{\dagger}}({\rm x})=\sqrt{\frac{2}{L_s}}\sum_{|k|<\Lambda_{\rm c}}\!\!\!\ee^{-\ii k{\rm x}}a_{k}^{\phantom{\dagger}}, \hspace{1ex} \psi_d^{\phantom{\dagger}}({\rm x})=\sqrt{\frac{2}{{L_s}}}\sum_{|k|<\Lambda_{\rm c}}\!\!\!\ee^{-\ii k{\rm x}}b_{k}^{\phantom{\dagger}},
\eeq
where $a_k^{\phantom{\dagger}},b_k^{\phantom{\dagger}}$ are momentum operators obtained from the odd- and even-site fermionic operators,  respectively. Therefore,  this long-wavelength approximation  focuses on  local aspects of the bands, and one might be loosing relevant information about  global topological features that  would require the  knowledge of the complete band structure.  Indeed, one finds that the Berry connection for  the lowest-energy band $ \mathcal{A}_-(k)=\bra{-\epsilon_k}\ii\partial_k\ket{-\epsilon_k}$ of the full SSH model~\eqref{eq:SSH} is
 \beq
 \mathcal{A}_-(k)=\frac{1-(1-\Delta)\cos ka}{2\big(1+(1-\Delta)^2-2(1-\Delta)\cos ka \big)}.
 \eeq
The ground-state of the SSH model at half filling $\ket{\rm gs}=\otimes_{k\in{\rm BZ}}\ket{-\epsilon_k}$ displays a polarization proportional to a non-trivial topological invariant~\cite{berry_rmp}: the so-called  {\it Zak's phase}~\cite{zak_phase}. This invariant is obtained by integrating the Berry connection over all the occupied momenta
\beq
\label{eq:Zak}
 \varphi_{{\rm Zak}}=\int_{\rm BZ}{\rm d}k\mathcal{A}_-(k)=\pi\big(\vartheta(\Delta)-\vartheta(2-\Delta)\big),
\eeq
where we have introduced  Heaviside's step function $\vartheta(x)=1$ for $x>0$, and zero otherwise. Therefore, this Zak's phase can be associated to a  {\it gauge-invariant topological Wilson loop} $W=\ee^{\ii\varphi_{{\rm Zak}}}$, which becomes non-trivial $W=-1$ when the  dimerization lies in $\Delta\in(0,2)$. This is precisely  the region where the SSH model hosts an SPT phase, a topological insulator in the  $\mathsf{BDI}$ symmetry class: the ground-state is characterized by  a non-vanishing topological invariant  respecting the    symmetries of the underlying Hamiltonian. These correspond to  time-reversal   $\mathsf{T}$ $\sigma^zh(-k)^*\sigma^z=h(k)$, particle-hole $\mathsf{C}$ $h(-k)^*=-h(k)$, and sub-lattice $\mathsf{S}$ $\sigma^zh(k)\sigma^z=-h(k)$ symmetry, such that $\mathsf{T}^2=\mathsf{C}^2=+1$~\cite{table_ti_1,*table_ti_2}. 

As announced above, in order to capture the correct topological features, one cannot na{\"i}vely restrict to long-wavelengths  $|k|<\Lambda_{\rm c}$~\eqref{eq:naive_continuum_limnit_PBC}, since the information about the topological mass $m_{\rm t}(k)$ at the borders of the Brillouin zone $|k-\pi/a|<\Lambda_{\rm c}$ is also important. In the following section, we use the bulk-boundary correspondence for such SPT phase to derive the correct long-wavelength approximation.

\subsection{SSH Model - Continuum limit}
\label{sec:edge_qed}

In this subsection we derive the correct long-wavelength approximation  that is valid for the non-interacting Schwinger model of Eq.~\eqref{eq:dimerised_KS_schwinger} in the main text. We build on the bulk-edge correspondence, which states that the non-vanishing bulk topological invariant  $\varphi_{{\rm Zak}}$ in Eq.~\eqref{eq:Zak} is related to the presence of robust zero-energy modes localized to the boundaries of the sample, the so-called {\it topological edge states}. Our goal now is to revisit the continuum limit in a way that these edge states appear naturally. 

Instead of considering periodic boundary conditions as in the previous subsections, we impose Dirichlet boundary  conditions for an open  finite chain. In the continuum limit, where $a\to 0$ and $N_s\to\infty$ with a fixed length $L_s=N_sa$, we can express the fermionic lattice operators as fields
$c_n\to\Psi({\rm x})=\sqrt{2/L_s}\sum_{k}\sin(k{\rm x})c_k,$ where  $k=\frac{\pi}{L_s}j$ and  $j\in\mathbb{N}$. Such fields fulfill  directly the  boundary conditions $\Psi(0)=\Psi(L_s)=0$.

\begin{figure}
\centering
  \includegraphics[width=1\columnwidth]{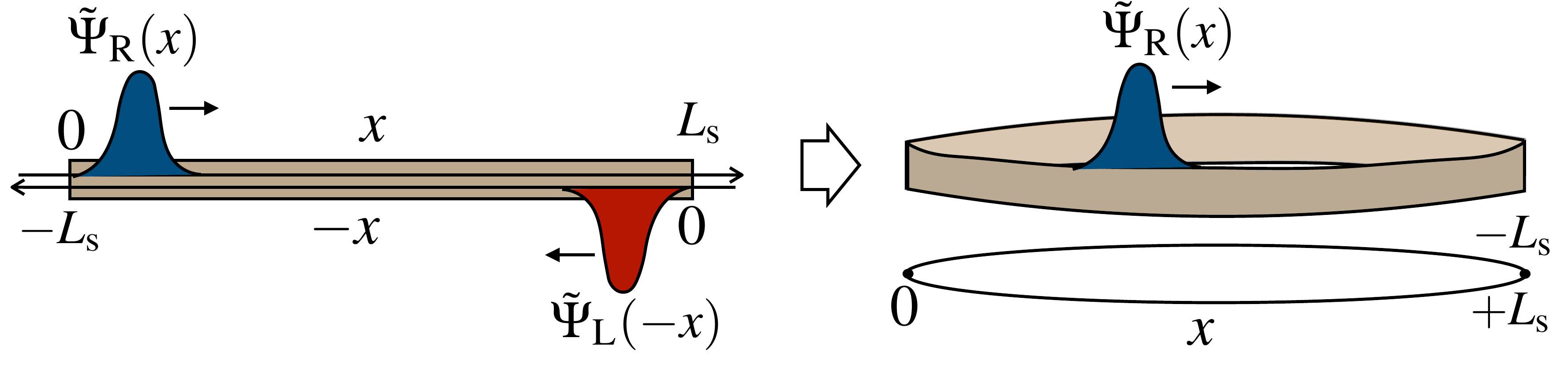}
  \caption{\label{Fig:modes_scheme} Chiral modes for Dirichlet boundary conditions: For a finite open chain, the right- and left-moving modes satisfying $\tilde{\Psi}_{\eta}(0)=\tilde{\Psi}_{\eta}(L_{\rm s})=0$, must fulfill  $\tilde{\Psi}_{\rm R}(x)=-\tilde{\Psi}_{\rm L}(x)$, such that one may study  modes of a fixed chirality living in an annulus, i.e. enlarged chain with periodic boundary conditions $\tilde{\Psi}_{\rm R}(L_{\rm s})=\tilde{\Psi}_{\rm R}(-L_{\rm s})$.}
\end{figure}

 In order to unveil the low-energy excitations that resemble  Dirac fermions, the standard approach in one-dimensional models  is to break the field operator into right- and left-moving components $\Psi({\rm x})=\ee^{\ii k_{\rm F}{\rm x}}\tilde{\Psi}_{\rm R}({\rm x})+\ee^{-\ii k_{\rm F}{\rm x}}\tilde{\Psi}_{\rm L}({\rm x})$, where  $\{\tilde{\Psi}_{\eta}({\rm x})\}_{\eta={\rm R,L}}$ are  slowly-varying envelopes that allow for a gradient expansion~\cite{affleck_les_houches}. For an open chain, however, these right- and left-moving fields are not independent, but  must instead fulfill $\tilde{\Psi}_{\rm L}(-{\rm x})=-\tilde{\Psi}_{\rm R}({\rm x})$ by imposing the   Dirichlet boundary conditions~\cite{fab_gog_open_boundaries} (see Fig.~\ref{Fig:modes_scheme}). Accordingly, the left-moving component can be obtained from the right-moving one, and one can focus on the right movers in a doubled chain with  periodic  conditions $\tilde{\Psi}_{\rm R}(-L_s)=\tilde{\Psi}_{\rm R}(+L_s)$.

In the present case,  we are interested in the universal properties of Eq.~\eqref{eq:dimerised_KS_schwinger} for $0<\Delta\ll 1$, which are obtained by making a long-wavelength   approximation around $k_{\rm F}=\pi/2a$ (i.e. wave-vector where the dispersion relation for open boundary conditions crosses the zero of energies). We can then restrict to momenta around the origin of the Brillouin zone  $|k-\pi/2a|<\Lambda_{\rm c}\ll1/a$, and perform a gradient expansion of the fermionic fields that yields a matter sector governed by
\beq
H_{\rm SSH}=\int_{0}^{L_s}\!\!{\rm dx}\!\!\sum_{\eta={\rm L,R}}\!\! s_\eta\left(\tilde{\Psi}_{\eta}^\dagger({\rm x})\ii\partial_{\rm x}\tilde{\Psi}_{\eta}^{\phantom{\dagger}}({\rm x})+\ii m\tilde{\Psi}_{\eta}^\dagger({\rm x})\tilde{\Psi}_{\bar{\eta}}^{\phantom{\dagger}}({\rm x})\right),
\eeq
where we have introduced $s_\eta=(1-2\delta_{\eta,{\rm R}})$,  $\bar{\eta}={\rm L,R}$ for ${\eta}={\rm R,L}$, and we recall that $m=-\Delta/a$. Here, the right- and left-moving fermions can be related to the original spinor components as follows $\tilde{\Psi}_{\rm R}=(\tilde{\Psi}_{\rm u}-\tilde{\Psi}_{\rm d})/\sqrt{2},  \tilde{\Psi}_{\rm L}=(\tilde{\Psi}_{\rm u}+\tilde{\Psi}_{\rm d})/\sqrt{2}$.

We can now use the condition $\tilde{\Psi}_{\rm L}(-{\rm x})=-\tilde{\Psi}_{\rm R}({\rm x})$ to get rid of the left-moving fields, and obtain the following continuum  field theory for the right movers $H_{\rm SSH}=\int_{-L_s}^{+L_s} {\rm dx}\mathcal{H}_{t{\rm D}}$, where 
\beq
\label{eq:non_local_mass}
\mathcal{H}_{t{\rm D}}=-\tilde{\Psi}_{R}^\dagger({\rm x})\ii\partial_{\rm x}\tilde{\Psi}_{R}^{\phantom{\dagger}}({\rm x})+\ii m\hspace{0.2ex}{\rm sgn}({\rm x})\tilde{\Psi}_{R}^\dagger({\rm x})\tilde{\Psi}_{R}^{\phantom{\dagger}}(-{\rm x}).
\eeq
Therefore, the na{\"i}ve continuum limit with massive Dirac fermions~\eqref{eq:naive_continuum_limnit_PBC}, must be replaced by this effective Hamiltonian field theory where the Dirac fermions  display a {\it non-local mass} that changes sign at $x=0$. This can be interpreted as a non-local version of the Jackiw-Rebbi quantum field theory, where fermionic zero-modes are localized within a kink excitation of a scalar field, which effectively changes the sign of the local  mass term~\cite{jackiw_rebbi}. In fact, this continuum field theory~\eqref{eq:non_local_mass} can be exactly diagonalized, and leads to  two types of solutions: {\it (i)} bulk energy levels with $\epsilon(k)=\pm \sqrt{m^2+k^2}$, where we recall that momentum is quantized $k=\pi j/L_s$ with $j\in\mathbb{N}$,    such that the solutions fulfill the Dirichlet boundary conditions.  Accordingly, these plane-wave solutions are delocalized within the bulk of the chain, and have a relativistic dispersion relation: they correspond to the previous massive Dirac field in the na{\"i}ve continuum limit~\eqref{eq:naive_continuum_limnit_PBC} . Additionally, in the thermodynamic limit, we find  {\it (ii)} a zero-energy mode localized at ${\rm x}=0$  with wave-function $\tilde{\chi}_0({\rm x})\approx C\ee^{-|{\rm x}|/\xi}$, where $\xi=a/\Delta \ll L_s\to\infty$ and  $C=\sqrt{\Delta a}$. Therefore, provided that $\Delta>0$ (otherwise the solution is not normalizable), we find a zero-mode exponentially localized to ${\rm x}=0$. This coincides  precisely with  the   topological edge state localized at the left boundary at ${\rm x}=0$, while the remaining edge state  at ${\rm x}=L_s$ can be recovered by means of inversion symmetry.

 After going back to the physical un-doubled chain, and introducing the fast-oscillating terms components to these envelopes, the zero-energy  solutions $\epsilon_{\rm L}=\epsilon_{\rm R}=0$  can be expressed as
\beq
\label{eq:sup_edge_states}
{\chi}_{\rm L}({\rm x})=C\ee^{-\frac{{\rm x}}{\xi}}\sin(k_{\rm F} {\rm x}),\hspace{1ex}{\chi}_{\rm R}({\rm x})= C\ee^{-\frac{(L_s-{\rm x})}{\xi}}\sin(k_{\rm F} (L_s-{\rm x})),
\eeq
which, in addition to the exponential decay  from the boundaries,  also show an oscillating character $\sin(\pi j/2)$ [$\sin(\pi (N-j)/2)$] such that the left-most [right-most] edge state only populates the even (odd) sites. As a consistency check, we note that this exponential decay and alternating behavior has been also found for the SSH model using completely different approaches (see e.g.~\cite{intro_ti}).

With these results, the na{\"i}ve continuum limit for the SSH model in Eq.~\eqref{eq:naive_continuum_limnit_PBC} gets superseded by $H_{\rm SSH}=\int_{0}^{L_s} {\rm dx}\mathcal{H}_{t{\rm D}}$, where 
 \beq
 \mathcal{H}_{t{\rm D}}=\overline{\Psi}({\rm x})(-\ii\gamma^1\partial_{\rm x}+\Delta/a){\Psi}({\rm x})+\sum_{\eta={\rm L,R}}\!\!\!\epsilon_{\eta}|\chi_{\eta}({\rm x})|^2\hat{\eta}^\dagger\hat{\eta},
 \eeq
 which is valid for for $0<\Delta\ll 1$ and corresponds to the Hamiltonian of Eq.~\eqref{eq:correct_continuum_ssh} of the main text.


%
\section{\texorpdfstring{$\mathbb{Z}_N$}{Zn}-topological Schwinger model on the lattice}\label{sec:ZnModelHamiltonian}
The goal of this section is to explain in more detail the Hamiltonian approach to lattice gauge theories for the discrete Abelian gauge group $\mathbb{Z}_{N}$, which gives access to the properties of compact QED in the large-$N$ limit \cite{hamiltonian_z_n_gauge}. For the massive Schwinger model of Eq.~\eqref{eq:schwinger_lagrangian} of the main text, this offers an alternative \cite{zn_presentation} to the Kogut-Susskind approach based on the Hamiltonian
\beq
\label{eq:Z_n_schwinger1}
\begin{split}
H_{m{\rm S}}^{\mathbb{Z}_N}\! =a\!\sum_{n=1}^{N_s} & \!\left(\!\frac{-1}{2a} \left(\ii c_n^{{\dagger}}\tilde{U}_{n}^{\phantom{\dagger}}c_{n+1}^{\phantom{\dagger}}+{\rm H.c.}\!\right)\!+m_{\rm s}(-1)^nc_n^{{\dagger}}c_n^{\phantom{\dagger}}\right.\\
&+\left.a(\tilde{V}_n+\tilde{V}_n^\dagger-2)\right)\!,
\end{split}
\eeq
where we have introduced two types of  unitary link operators $\tilde{U}_n,\tilde{V}_n$ that obey the $\mathbb{Z}_N$  algebra.
Accordingly, instead of using the rotor-angle operators of the Kogut-Susskind approach, one uses link operators  fulfilling $\tilde{U}_{n}^N=\tilde{V}_{n}^N=\mathbb{I}$, and $\tilde{V}^\dagger_{n}\tilde{U}_n\tilde{V}_{n}=\ee^{\ii2\pi/N}\tilde{U}_n$. In analogy to the Kogut-Susskind approach, using the electric-flux eigenbasis $\tilde{V}_n\ket{v}=v\ket{v}$ with $v\in\mathbb{Z}_N$, the remaining link operators act as ladder operators that raise the electric flux by one quantum $\tilde{U}_n\ket{v}=\ket{v+1}$. The main difference is that, in contrast to the  Kogut-Susskind approach,   the ladder operators have  a cyclic constraint $\tilde{U}_n\ket{N}=\ket{1}$.

We note that these link operators can be defined in terms of the vector potential  and the electric field  $\tilde{U}_n={\rm exp}\{\ii a g A_n\}$,   $\tilde{V}_n={\rm exp}\{\ii \frac{2\pi}{ N}\frac{E_n}{g}\}$. In this way, the $\mathbb{Z}_N$ algebra $[\tilde{U}_n,\tilde{V}_n]=\ee^{\ii 2\pi/N}$ can be  satisfied by imposing the usual  canonical commutation relations $[E_n,A_m]=\ii\delta_{n,m}/a$, which have  the correct continuum limit $[E(x),A(y)]=\ii\delta(x-y)$. Note also that the gauge-group condition $\tilde{U}_n^N=\tilde{V}_n^N=\mathbb{I}$ requires that the electric-flux eigenvalues of $\tilde{L}_n=E_n/g$ should span $\sigma(\tilde{L}_n)=\{-\half(N-1),\cdots,\half(N-1)\}$. This yields $\sigma(\tilde{L}_n)\to \mathbb{Z}$ in the large-$N$ limit, which corresponds to the spectrum of the rotor operator $L_n$ of the Kogut-Susskind approach. In the same manner, the eigenvalues of the vector potential should lie in $\sigma(ag A_n)=\{-\pi(N-1)/N,\cdots,\pi(N-1)/{N}\}\to[-\pi,\pi]$, corresponding to the basis of the angle operator  $\Theta_n$ in the Kogut-Susskind approach, and leading to compact QED$_2$. We remark that, as emphasized in~\cite{zn_presentation}, the electric-energy term in Eq.~\eqref{eq:Z_n_schwinger1} can be substituted by an arbitrary  function $(V_n+V_n^\dagger-2)\to f(V_n)=f^\dagger(V_n)$, and we will focus on $ f(V_n)=\half g^2\tilde{L}_n^2$~\cite{alternative_z_n_gauge, zn_study}.

As shown in~\cite{zn_study}, the properties of the massive Schwinger mode with vacuum angle  $\theta=\pi$ can be recovered from a large-$N$ scaling  of the  $\mathbb{Z}_{N}$ massive Schwinger model~\eqref{eq:Z_n_schwinger1}. 

By introducing our alternative discretization presented in the Letter, we arrive to the lattice Hamiltonian of Eq.~\eqref{eq:dimerised_KS_schwinger} of the main text, i.e.
\beq
\label{eq:dimerised_KS_schwinger_Z_N}
H^{\mathbb{Z}_N}_{t{\rm S}}\!=a\!\sum_{n=1}^{N_s} \!\left(\!\frac{-1}{a} \left(\ii (1-\delta_n)c_n^{{\dagger}}\tilde{U}_{n}^{\phantom{\dagger}}c_{n+1}^{\phantom{\dagger}}+{\rm H.c.}\!\right)\!+\frac{g^2}{2}\tilde{L}_n^2\right)\!,
\eeq
where $\delta_{2n}=0$ and $\delta_{2n-1}=\Delta$. 

In order to take into account  Gauss's law,  we also introduce the operator 
\beq
\label{eq:magnifico_gauss_law}
G_{n}=c_{n}^{\dagger}c_{n} + \frac{1}{2a}[(-1)^{n}-1]-\frac{1}{a}(\tilde{L}_{n}-\tilde{L}_{n-1}).
\eeq
Accordingly,  $\left |  \psi \right > $ is a physical state if it satisfies the condition 
$
G_{n}\left |  \psi \right > = 0 \; \forall n 
$.
This is a very important constraint that allows us to construct the physical Hilbert space of the topological $\mathbb{Z}_{N}$ model for numerical simulations by a DMRG algorithm.

\section{Critical lines: scaling analysis}\label{sec:critical_lines_scaling}

In this Section we show the finite-size scaling analysis for determining the exact location of the critical lines separating the SPT, the confined and the fermionic condensate phases in the phase diagram of the discretization of the Schwinger model we introduced. 

\begin{figure}
 \centering
  \includegraphics[width=0.8\columnwidth]{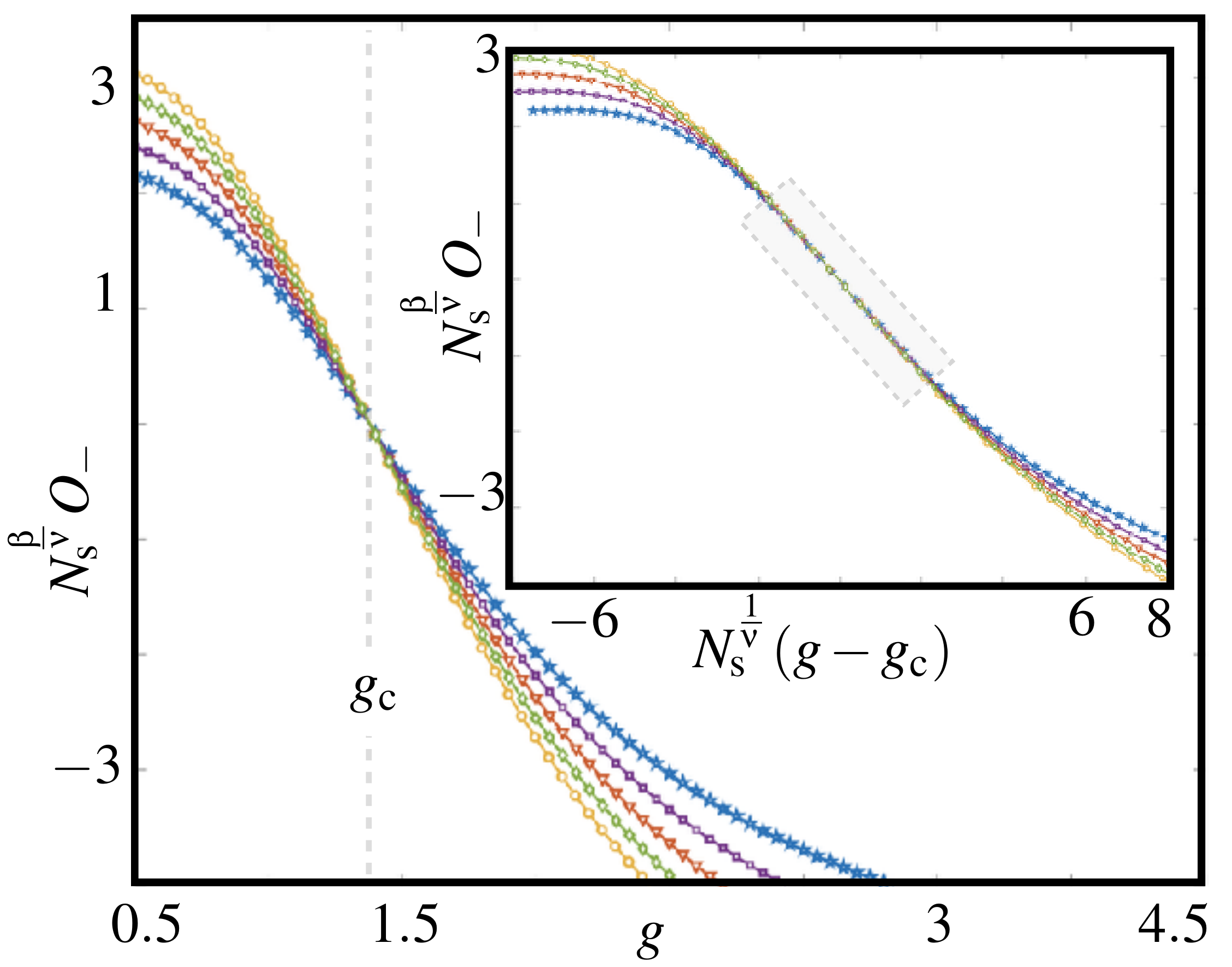}
  \caption{\label{Fig:scaling_topo1} Finite-size scaling for the topological correlator of the $\mathbb{Z}_3$ topological Schwinger model.  (main panel) Scaling quantity $N_{\rm s}^{\beta/\nu}O_{-}$  for the topological correlator~\eqref{eq:magnifico_topological_order_quantity_Supplementary} calculated for $\Delta = 0.5$ as a function of the gauge coupling for various system sizes $N_{\rm s}\in\{24,28,32,36,40\}$ (top to bottom). The crossing point of all curves yields a value of the critical point separating the SPT and C phases of $g_{\rm c} \approx 1.384$. (inset) Universal scaling of the topological correlator within the Ising universality class $\nu=1$ and $\beta=1/8$, i.e. data collapse of the curves $N_{\rm s}^{\beta/\nu}O_{-}$ as function of $N_{\rm s}^{1/\nu}(g-g_{\rm c})$ displayed in the shaded region.}
\end{figure}

\subsection{Finite-size scaling: topological order parameter}

The critical line separating the SPT from the trivial confined phase in the thermodynamic limit can be determined by a finite-size scaling of a topological order parameter $O_-$ recently introduced in Ref.~\cite{ssh_order_parameters} for the SSH model and defined as $O_{-} =\frac{2}{N_{\rm s}}\sum_{n=1}^{N_{\rm s}/2}\bra{\rm gs} O^{(2n-1)}_{-}\ket{\rm gs}$ with
\beq
\label{eq:magnifico_topological_order_quantity_Supplementary}
O^{(j)}_{-} = \frac{3}{2} \left(c_{j}^{\dagger}c_{j+1}^{\phantom{\dagger}} +  c_{j+1}^{\dagger}c_{j}^{\phantom{\dagger}}  \right) + \rho_{j}\rho_{j+1} -  \frac{1}{2}(\rho_{j}+\rho_{j+1}),
\eeq
where  $\rho_{j}=c^\dagger_jc^{\phantom{\dagger}}_j$  are fermion density operators. Throughout all this Section we will set the lattice spacing $a=1$.

Finite-size scaling theory predicts that there exist a universal function $\lambda(x)$ and two critical exponents $\nu$ and $\beta$ such that the quantity $O_{-}$ will behave as
\beq
\label{eq:magnifico_topological_scaling_relation}
N_{\rm s}^{{\beta}/{\nu}}O_{-} = \lambda \left ( N_{\rm s}^{{1}/{\nu}} (g-g_{\rm c}) \right)
\eeq
for coupling $g$ close enough to the critical point $g_c$. Since in the SSH model, the quantum phase transition between the topological and the trivial phase has critical exponents $\beta=1/8$ and $\nu=1$, we assumed these values in Eq.~\eqref{eq:magnifico_topological_scaling_relation}.

We note that for $g=g_{\rm c}$, the value  $\lambda(0)$ and thus the value $N_{\rm s}^{{\beta}/{\nu}}O_{-}$, because of Eq.~\eqref{eq:magnifico_topological_scaling_relation}, become independent of the system size and one expects to find a crossing of the curves representing $N_{\rm s}^{{\beta}/{\nu}}O_{-}$ for different $N_{\rm s}$ precisely at the critical point. Therefore, after fixing $\Delta$, we determined $g_c$ by plotting the l.h.s of Eq.~\eqref{eq:magnifico_topological_scaling_relation} as a function of $g$  and for different system sizes and by looking at the point $g_c$ where the curves intersected.

We computed $O_-$ by using our DMRG algorithm for open boundary conditions, where we keep $m = 1000$ states in the iterative diagonalization and coarse graining of a lattice of different sizes (up to $N_s=80$ sites).

The main panel of Fig.~\ref{Fig:scaling_topo1} shows examples of the quantity $N_{\rm s}^{\beta/\nu}O_{-}$ as a function of $g$ for different values of $N_{\rm s}$ and $\Delta=0.5$. It is possible to see that the crossing of the curves allows us to predict a critical point at  $g_{\rm c} \approx 1.384/a$. Morever, to check the initial hypothesis concerning the values of the critical exponents $\beta=1/8$ and $\nu=1$, we  analyze the quantity $N_{\rm s}^{{\beta/\nu}}O_{-}$ as a function of the argument $N_{\rm s}^{1/\nu} (g-g_{\rm c})$. In this case, for different system sizes, we should observe a universal behavior when $g \approx g_{\rm c}$ (i.e. a collapse of the different curves into a single one). This is exactly what is shown in the inset of Fig.~\ref{Fig:scaling_topo1}, confirming in this way the initial hypothesis about the universality class of the SPT-C phase transition. 

In the same spirit, we can now fix a particular value of $g$, and calculate the topological order parameter by varying the dimerization  $\Delta$ to compute the critical dimerization $\Delta^{O_-}_c$.  By varying $g$ we  determined the critical points related to the transition SPT-C. The resulting values are shown Table~\ref{tab:table_of_critical_values_vertical1}. 

As can be observed in the last row of this table, when the gauge coupling $g$ is sufficiently large the topologica SPT-C transition is absent. This means that the SPT phase disappears for large $g$, as conjectured in the Letter.

\begin{figure}
\centering
\includegraphics[width=0.8\columnwidth]{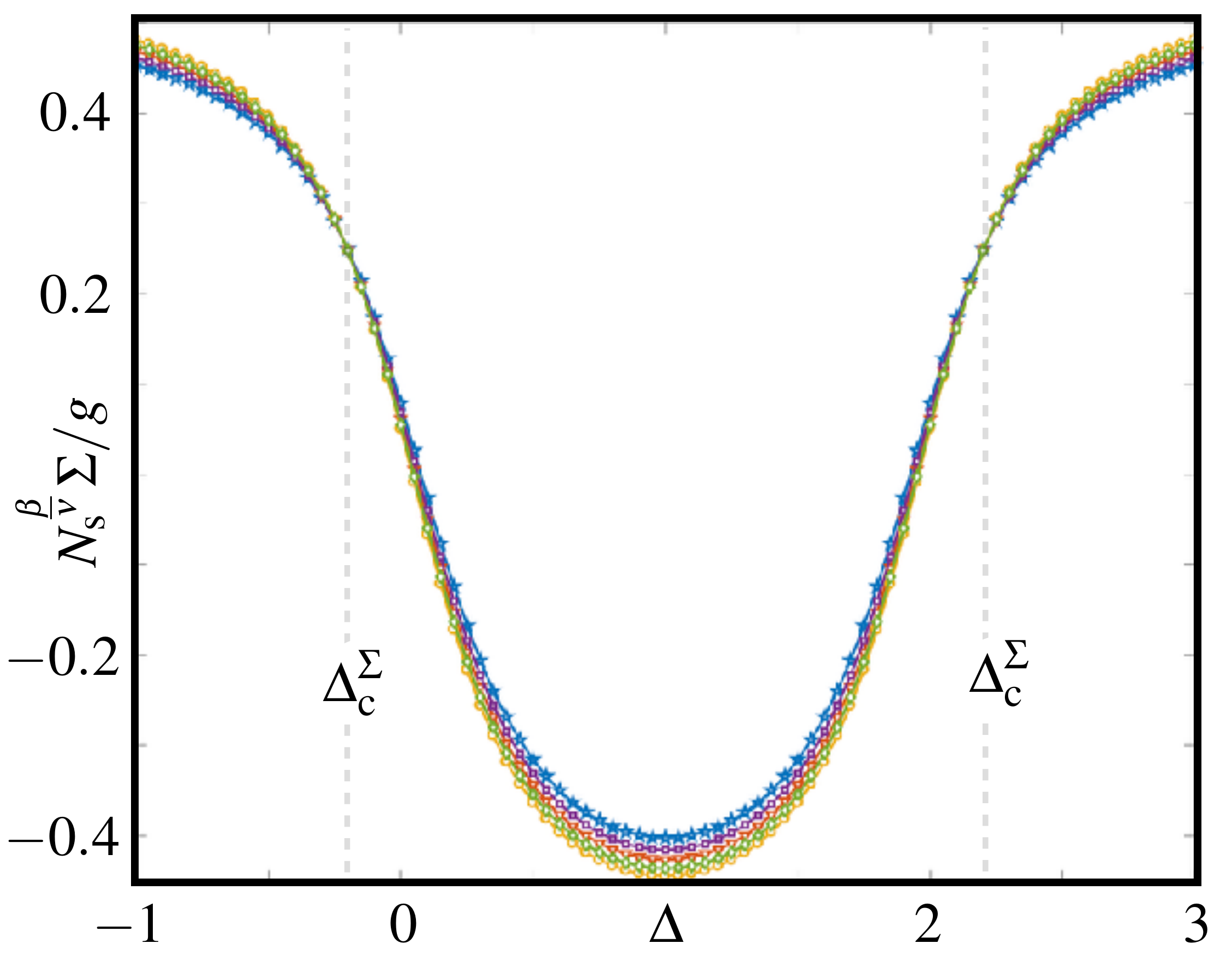}
\caption{\label{Fig:scaling_electric1} Finite-size scaling for the electric order parameter  of the $\mathbb{Z}_3$ topological Schwinger model: Scaling quantity $N_{\rm s}^{\beta/\nu}\Sigma/g$  for the electric-field order parameter $\Sigma$, calculated for $g = 0.6$ as a function of the dimerization for various system sizes $N_{\rm s}\in\{24,28,32,36,40\}$ (top to bottom). The two crossing points of all curves yield the values of the critical dimerizations of $\Delta_{\rm c}^{\Sigma} \approx -0.215$ and $\Delta_{\rm c}^{\Sigma} \approx 2.216$ that are symmetric with respect to the line $\Delta = 1$.}
\end{figure}

\subsection{Finite-size scaling: electric order parameter}


As conjectured in the Letter, the phase transition  separating the confined phase (C) from the fermion condensate (FC) should be analogous to the one of the standard massive Schwinger model. In the context of the $\mathbb{Z}_{N}$ approach~\cite{zn_study}, such a transition  can be  detected  numerically by  the electric field order parameter $\Sigma = \sum_{n=1}^{N_{s}} \bra{\rm gs} E_{n}  \ket{\rm gs}/{N_s}$.

Following the scheme of the previous transition SPT-C, we can compute the critical dimerization $\Delta^\Sigma_{\rm c}$ by performing a finite-size scaling analysis  with
\beq
\label{eq:magnifico_electric_scaling_relation}
N_{\rm s}^{{\beta}/{\nu}}\Sigma = \lambda \left ( N_{\rm s}^{{1}/{\nu}} (\Delta-\Delta^\Sigma_{\rm c}) \right ),
\eeq
where we use the critical exponent of the universality class of the massive Schwinger model, which is the 2D Ising class $\beta=1/8$, $\nu=1$~\cite{massive_schwinger_theta,lattice_schwinger_critical_point,schwinger_dmrg}.

Examples of the quantity  $N_{\rm s}^{{\beta}/{\nu}}\Sigma$ are plotted in Fig.~\ref{Fig:scaling_electric1} as a function of $\Delta$ for different system sizes $N_{\rm s}$ and for $g=0.6$. The two critical points $\Delta_{\rm c}^{\Sigma} \approx -0.215$ and $\Delta_{\rm c}^{\Sigma} \approx 2.216$ (symmetrical with respect to the value $\Delta=1$ as expected) are found at the crossing point of all the lines.

Thus, the behavior of $N_{\rm s}^{{\beta}/{\nu}}\Sigma$ when varying $g$ can be used to determine the critical points $\Delta_{\rm c}^{\Sigma}$ related to the transition C-FC. The resulting values are shown in Table~\ref{tab:table_of_critical_values_vertical1}. 


\begin{table}
\centering
\newcolumntype{C}{>{$}c<{$}}
\begin{tabular}{|C|C|C|C|C|}
\hline 
g & \Delta_{c}^{O_{-}} & \Delta_{c}^{O_{-}} & \Delta_{c}^{\Sigma} & \Delta_{c}^{\Sigma}  \tabularnewline
\hline 
\hline 
0.01            &     \phantom{-}0.009            &               1.994            &     \phantom{-}0.008            &               1.995   \\  \hline
0.05            &               -0.022            &               2.028            &               -0.024            &               2.030   \\  \hline
0.10            &               -0.062            &               2.069            &               -0.064            &               2.072   \\  \hline
0.20            &               -0.120            &               2.118            &               -0.121            &               2.119   \\  \hline
0.60            &               -0.162            &               2.160            &               -0.215            &               2.216   \\  \hline
1.00            &               -0.051            &               2.052            &               -0.257            &               2.259   \\  \hline
3.00            &               //                &               //               &               -0.210            &               2.211   \\  \hline
\end{tabular}\caption{\label{tab:table_of_critical_values_vertical1} Critical values of $\Delta$ (related to the two transitions FC-C and SPT-C) obtained for different values of $g$. The numerical error is equal to $10^{-3}$.}
\end{table}


\section{Wave functions of the edge modes}\label{sec:edge_mode_wf}

In this Section we explain how to compute the wave functions of the two zero-energy edge modes by DMRG simulation.

Let us start by considering  the SPT phase in the non-interacting limit $g=0$, where two zero-energy edge modes are present. Since the Hamiltonian commutes with the number operator, the Hilbert space can be divided into sectors with a fixed number of particles. Neglecting small finite-size corrections to their energies, which will eventually  disappear in the thermodynamic limit, there will be four degenerate states in the ground-state manifold:  $|\text{gs}_{N_{\rm s}-1}\rangle$ in the sector with $N_{\rm s} - 1$ particles;  $|\Phi_\text{L}\rangle$ and $|\Phi_\text{R}\rangle$ in the sector with $N_s$ particles with the leftmost or rightmost edge modes populated;  $|\Phi_\text{L} ,\Phi_\text{R}\rangle$  in the sector with $N_s+1$ particles hosting both populated  edge modes. Let now $\Phi_\text{L}^\dag$ represent the operator that excites the leftmost zero-energy mode, i.e. $\Phi^{\dagger}_{\rm L}=\sum_n\alpha^nc_n^{\dagger}$ for some $\alpha<1$, and analogously for the rightmost zero-energy mode $\Phi^{\dagger}_{\rm R}$. Accordingly, we can obtain the ground-state with $N_{\rm s}$ particles as $|\Phi_{\rm L} \rangle = \Phi^{\dagger}_{\rm L}\ket{{\rm gs}_{N_{\rm s}-1}}$ (or equivalently $|\Phi_{\rm R} \rangle = \Phi^{\dagger}_{\rm R}\ket{{\rm gs}_{N_{\rm s}-1}}$), and the ground state with $N_{\rm s}+1$ particles as $| \Phi_{\rm L}, \Phi_{\rm R} \rangle = \Phi^{\dagger}_{\rm L}\Phi^{\dagger}_{\rm R}\ket{{\rm gs}_{N_{\rm s}-1}}$. Using the DMRG algorithm, we can numerically target the lowest energy state in sectors with a generic number of particles, and we can thus calculate the following expectation value
\beq
\label{eq:edge_obs1}
B_{n} = \langle \Phi_{\rm L}, \Phi_{\rm R} | c^{\dagger}_{n}c_{n} | \Phi_{\rm L}, \Phi_{\rm R}  \rangle. 
\eeq
Note that in the non-interacting limit, by applying Wick's theorem, this observable becomes 
\beq
\begin{split}
B_{n} & = |\bra{\rm gs_{N_{\rm s}-1}}c_{n} | \Phi_{\rm L} \rangle |^{2} \\ & + |\bra{\rm gs_{N_{\rm s}-1}}c_{n} | \Phi_{\rm R} \rangle |^{2} \\& +\bra{{\rm gs}_{N_{\rm s}-1}} c^{\dagger}_{n}c_{n}\ket{{\rm gs}_{N_{\rm s}-1}}. 
\end{split}
\eeq
Interestingly, given the above expression of the edge operators, the first two terms of the above expression contain the 
 the probabilities associated to the edge-state wave-functions. These wave-functions can thus be obtained  by calculating numerically
\beq
\label{eq:edge_prob}
\psi_{n}^{2} = B_{n} -  \langle {\rm gs}_{N-1}  | c^{\dagger}_{n}c_{n}| {\rm gs}_{N-1} \rangle.
\eeq
Recalling that the operator $\Phi_{L}$ ($\Phi_{R}$) has support only on even (odd) sites \eqref{eq:sup_edge_states}, it is possible to reconstruct the amplitude of the left-most (right-most) edge mode by plotting the quantity $\psi^{2}_{n}$ as a function of even (odd) $n$. 

We expect that the behavior of the observable~\eqref{eq:edge_obs1} will hold in the interacting regime, giving us a method to compute the {\it many-body zero-energy edge modes} of the topological Schwinger model, as reported in the plots of Fig.~\ref{Fig:o_param_1}(d).


\section{Robustness of the SPT phase in the large-\texorpdfstring{$N$}{N} limit}\label{sec:robustness_other_n}

In this work we studied the $\mathbb{Z}_3$ topological Schwinger model, leaving to a follow-up study the complete analysis of the phase diagram of the models with $N>3$. 

\begin{figure}
\centering
\includegraphics[width=0.8\columnwidth]{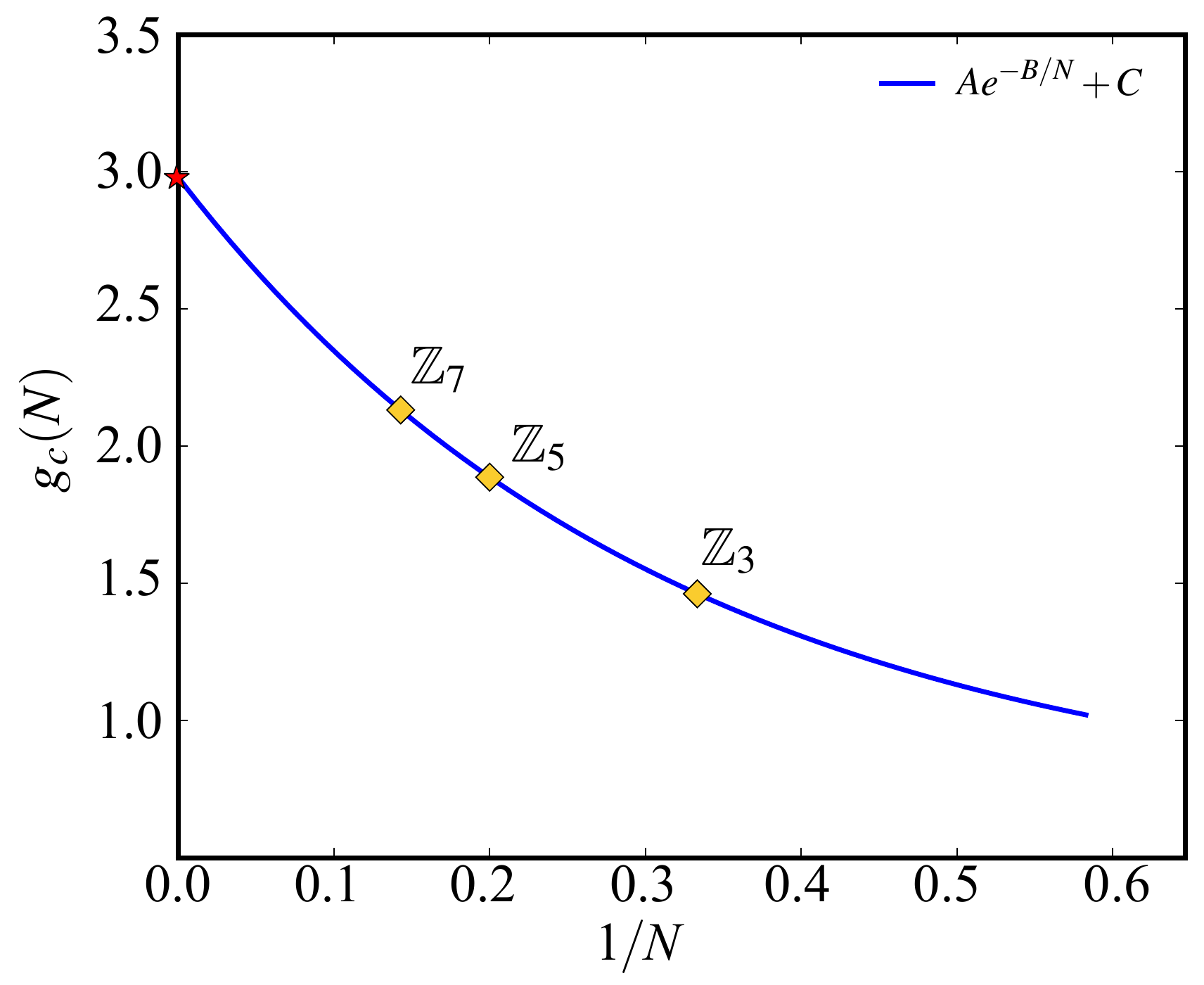}
\caption{\label{Fig:fit_gn_lin}  Critical points $g_c(N)$ (yellow squares) separating the SPT from the confined phases for the topological Schwinger models implemented by the $\mathbb{Z}_N$ approach with $N=3,5,7$ algebras. The solid line represents the function $Ae^{-B/N}+C$ that fits the points and gives a finite value of $g_c(\infty) \approx 2.979$ (marked with a red star) in the limit $N\rightarrow\infty$ indicating the stability of the SPT phase.}
\end{figure}

However, in order to understand the robustness of the SPT phase when the link operators belong to different $\mathbb{Z}_N$ algebras we also computed the critical points $g_{c}$ related to the transition SPT-C for the particular line $\Delta = 1$ for the $\mathbb{Z}_{5}$ and $\mathbb{Z}_{7}$ models. We noticed that the critical value $g_{c}(N)$ grows with $N$ and the points can be fit with an exponential function of the form $g_{c}(N)=Ae^{-B/N}+C$, as shown in Fig.~\ref{Fig:fit_gn_lin}. Thus, the critical point $g_{c}(N)$ for $\Delta=1$ approaches a finite value in the $N\rightarrow \infty $ limit given by $g_{c}(\infty) \approx 2.979$. This shows that the SPT phase has a finite region of stability, in accordance to the analytical results obtained for the $U (1)$ topological Schwinger model.

\bibliography{biblio_schwinger_all}

\end{document}